\documentclass[a4paper]{article}

\usepackage{INTERSPEECH2020}
\usepackage{enumitem}
\usepackage{paralist}
\usepackage{xcolor}
\usepackage{mathrsfs}
\usepackage{bbm}
\usepackage{xcolor}
\definecolor{darkgreen}{rgb}{0.0, 0.2, 0.13}
\usepackage{url}
\usepackage{epstopdf}
\usepackage{hyperref}

\title{
TinyLSTMs: Efficient Neural Speech Enhancement for Hearing Aids
}
\name{Igor Fedorov$^{\ast,1}$\thanks{Symbol $^*$ denotes equal contribution. Corresponding authors: I. F. (igor.fedorov$@$arm.com), M.S. 
(\href{mailto:marko_stamenovic@bose.com}{marko{\textunderscore}stamenovic$@$bose.com})
}, {Marko Stamenovic}$^{\ast,2}$, {Carl Jensen}$^{\ast,2}$, {Li-Chia Yang}$^{\ast,2}$, {Ari Mandell}$^2$, {Yiming Gan}$^3$, Matthew Mattina$^1$, Paul N. Whatmough$^1$}
\address{
  $^1$Arm ML Research Lab, Boston, MA\\
  $^2${Bose Corp., Boston, MA} \\
  $^3${University of Rochester, Rochester, NY} 
  }
\email{}

\begin{document}

\maketitle
\begin{abstract}
Modern speech enhancement algorithms achieve remarkable noise suppression by means of large recurrent neural networks (RNNs). However, large RNNs limit practical deployment in hearing aid hardware (HW) form-factors, which are battery powered and run on resource-constrained microcontroller units (MCUs) with limited memory capacity and compute capability. In this work, we use model compression techniques to bridge this gap. We define the constraints imposed on the RNN by the HW and describe a method to satisfy them. Although model compression techniques are an active area of research, we are the first to demonstrate their efficacy for RNN speech enhancement, using pruning and integer quantization of weights/activations. We also demonstrate state update skipping, which reduces the computational load. Finally, we conduct a perceptual evaluation of the compressed models to verify audio quality on human raters. Results show a reduction in model size and operations of 11.9$\times$ and 2.9$\times$, respectively, over the baseline for compressed models, without a statistical difference in listening preference and only exhibiting a loss of 0.55dB SDR. Our model achieves a {computational} latency of 2.39ms, well within the 10ms target and 351$\times$ better than previous work.

\end{abstract}


\section{Introduction}
    The healthy ear is a complex, non-linear system, capable of operating over a large dynamic range. When the ear is damaged, the auditory system can be augmented with a hearing aid (HA), which performs some of the amplification and filtering functions that the ear is no longer able to do. Speech enhancement (SE) could ease listening difficulty in noisy environments, which ranks among the top concerns of HA users \cite{kochkin2000marketrak, abrams2015introduction, hearingtracker}.

\begin{table*}[ht]
\centering
\begin{tabular}{l c c}
\toprule
  Constraint & Specification & Rationale \\
\midrule    
Compute Complexity & {$\leq 1.55$ MOps/inf} & Compute budget to achieve {$\le$10ms} latency on target MCU \\
Model Size & $\leq 0.5$ MB & $0.5$ MB device Flash memory \\
Working Memory & $\leq 320$ KB & $320$ KB device SRAM \\
Data Type & Integer & Lowest energy on target MCU \\
Perceptual Quality & Minimal degradation vs. uncompressed model &  Compression should not affect user preference \\ 
\bottomrule
\end{tabular}
\caption{Model constraints. MOps/inf denotes 10$^6$ operations per frame inference. Target MCU is STM32F746VE~\cite{stm32f746ve}.
}
\vspace{-2em}
\label{table:constraints}
\end{table*}

Recent SE approaches are often embodied by recurrent neural networks (RNNs) \cite{Takeuchi2020,wilson2018exploring}. 
The SE model must achieve low audio latency to ensure listener comfort. Audio latency is defined as the delay between noise arriving at the HA and clean sound being produced for the listener. The amount of latency that can be tolerated varies depending on the HA type and how the user's own voice is processed \cite{stone1999delay,stone2002delay,stone2003delay}. Using previous work~\cite{stone1999delay,stone2002delay,stone2003delay} as a guideline, we target a maximum audio latency of 30ms. For the frame-based approach we employ, with 50\% overlap between frames and a causal model, the compute latency constraint for processing each frame is 10ms.

The HA form factor imposes another set of constraints, especially in combination with the frame processing requirement. Due to their small form factor, HAs use a microcontroller unit (MCU) hardware (HW) platform. MCUs enable cheap, low power computation, at the cost of severe memory and computation constraints \cite{fedorov2019sparse}. The MCU Flash memory restricts the maximum allowable model size (MS), whereas the on-chip SRAM memory upper bounds model working memory (WM), i.e. the memory used to store intermediate results. 
To achieve efficient computation, the SE model must be quantized to integer data types and we must minimize the number of operations (ops) required per second (ops/s), where an op represents a single addition or multiplication.
In this work, we target the STM32F746VE MCU~\cite{stm32f746ve} as a typical HW platform, which contains a 216MHz Arm Cortex-M7~\cite{arm-m7} with 512KB Flash memory, and 320KB SRAM. We used Mbed OS~\cite{arm-mbed} and CMSIS kernels~\cite{arm-cmsis,lai18-cmsis}. Table \ref{table:constraints} summarizes the SE model constraints.



A few recent works have considered similar constraints.
For example, Wilson et al. \cite{wilson2018exploring} use a black box optimizer to search for SE models in a family of causal and non-causal models that include compute-heavy convolutions on the model inputs. Model complexity is not explicitly constrained in the search, with the reported models in the $3.7-248$ MB range, violating the MS constraint. Moreover, some of the models include many layers of dilated convolutions on the front-end, which require roughly $4.4$ MB of WM, violating the WM constraint. 

Other works seek to prune \cite{wen2018learning} and quantize \cite{hou2019normalization} RNNs, but do not apply their techniques to SE. Although parameters are quantized in \cite{hou2019normalization}, activations are not and the resulting computation is not suitable for
integer arithmetic. Moreover, it is unclear from \cite{wen2018learning,hou2019normalization} if pruning and quantization can be jointly applied to RNNs. In Wu et al. \cite{wu2019}, a non-recurrent convolutional SE model is pruned and quantized. Their use of non-uniform quantization, however, requires
non-standard HW support~\cite{han2016eie} to avoid incurring 
non-trivial performance overhead from decoding 
each weight after loading it from memory. With a large receptive field, a convolutional model may also need large buffers operating at the audio sample rate.
This greatly inflates WM and dramatically tightens the constraints on computational time. Finally, Hsu et al.~\cite{hsu2018expquant} investigated separately quantizing the floating point mantissa and exponent values of recurrent and convolutional SE models \cite{hsu2018expquant}, but these quantized weights would still need to run in floating point HW and incur overheads for decompression. 

In this work, we present a methodology to generate optimized RNN SE models which meet the requirements in Table~\ref{table:constraints}. 
Firstly, we demonstrate pruning of SE LSTMs to reduce MS, WM and ops without incurring SE performance degradation. Extending \cite{wen2018learning}, we directly learn the pruning thresholds within the optimization, obviating costly hyperparameter search, resulting in 
$255\times$ less GPU hours (GPUH) compared to previous work \cite{wilson2018exploring}.
Secondly, we show for the first time that standard weight and activation quantization techniques extend well to SE RNNs.
Moreover, we show that pruning and quantization can be applied to SE RNNs jointly, which is also unique to our work. 
Finally, we propose a scheme for skipping RNN state updates, reducing the average number of operations. 

Our optimized SE models are evaluated using traditional objective metrics, as well as a subjective perceptual evaluation of the audio output.
The audio source files we used for this are available  online\footnote{\url{https://github.com/BoseCorp/efficient-neural-speech-enhancement}}. 
Our perceptual study is a significant improvement over \cite{Takeuchi2020,wilson2018exploring,wu2019,hsu2018expquant,erdogan2015phase,weninger2015speech}, because (compressed) SE models often exhibit acoustic artifacts not reflected in objective metrics such as SNR. Finally, we profile our models on an MCU to validate that they meet HW constraints, as shown in Table~\ref{table:constraints}.



\section{Background}
Let lowercase and uppercase symbols denote vectors and matrices, respectively, and let $X = \begin{bmatrix} x^1 & \cdots & x^N \end{bmatrix} \in \mathbb{R}^{M \times N}$.

\subsection{Speech Enhancement}
Let $x \in \mathbb{R}^N$ denote $N$ samples of a single channel time-domain speech signal. In SE, $x$ is corrupted by noise $v$ and the goal is to extract $x$ from $y = x + v$. In this work, denoising is applied in the time-frequency domain, whereby $y$ is transformed into $Y \in \mathbb{C}^{B_f \times B_t}$ using a short time Fourier transform 
where $B_t$ is the number of frames that $N$ is decomposed into and $B_f$ is the number of frequency bins. In this work, the denoiser is a mask $M \in \mathbb{R}_+^{B_f \times B_t}$ applied to the
spectrogram, such that the approximation of the target is given by $\hat{X} = M \odot \left \vert Y \right \vert \exp{\left(\measuredangle Y \right)}$, where $\odot$ denotes the Hadamard product and $\measuredangle Y$ is the phase of the noisy input \cite{wang2014training}. The mask is a function of $Y$ and learnable parameters $\theta$, i.e. $M = f_\theta(Y)$. Specifically, $f_\theta(\cdot)$ is a neural network whose parameters are learned by minimizing a phase-sensitive spectral approximation loss \cite{erdogan2015phase}:
\begin{align}
    {L}(\theta) = \left \Vert \lvert X \rvert^{0.3} - \lvert \hat{X} \rvert^{0.3} \right \Vert_F^2 + 0.113  \left \Vert X^{0.3} - \hat{X}^{0.3} \right \Vert_F^2 \label{eq:empirical risk},
\end{align}
where frames are power-law compressed with an exponent of $0.3$ to reduce the dominance of large values.

\subsection{Baseline Model Architecture}
\label{sec:proposed architecture}
Due to the latency requirement, we restrict our attention to causal models \cite{Takeuchi2020}, whereby $m^t = f_\theta \left(\begin{bmatrix} y^1 & \cdots & y^t \end{bmatrix} \right)$. We use an architecture consisting of a series of recurrent layers followed by fully connected (FC) layers. The recurrent layers serve to model interactions across time. For a recurrent layer, we use long-short term memory (LSTM) cells, which are stateful and have update rules $i^t = \sigma ( W_{xi} x^t + h^{t-1} W_{hi} + b_i ), r^t = \sigma ( W_{xr} x^t +  h^{t-1} W_{hr} + b_r ), o^t = \sigma ( W_{xo} x^t + h^{t-1} W_{ho} + b_o ), u^t = \tanh ( W_{xu} x^t + h^{t-1} W_{hu} + b_u )$,
\begin{align}
    c^t &= r^t \odot c^{t-1} + i^t \odot u^t, \; \; h^t = o^t \odot \tanh \left( c^t \right), \label{eq:lstm 5}
\end{align}
where $\sigma$ is the sigmoid function \cite{hochreiter1997long}. The baseline architecture consists of 2 unidirectional LSTM layers with 256 units each and 2 FC layers with 128 units each, with batch normalization between the last LSTM and first FC layers.
ReLU activation is applied after the first FC layer and sigmoid after the second. In all cases, the spectral input to the network is mapped onto a 128-dimensional mel space \cite{mel} and power-law compressed with an exponent of 0.3. The network output, which shares the dimensionality of the input, is inverted using the corresponding transposed mel matrix to produce spectral mask $M$.





\section{Optimizing LSTMs for HA Hardware}
\label{sec:optimizing lstms}
This section introduces optimizations for SE models, such as those in Section \ref{sec:proposed architecture}, to satisfy the constraints given in Table \ref{table:constraints}. We begin by delineating the dependence of MS and computational cost on the properties of the model. Then, in Sections \ref{sec:quantization}-\ref{sec:skip rnn} we describe our proposed approach.

\begin{table*}
\centering
\resizebox{\linewidth}{!}{%
\begin{tabular}{l c c c c c c c c c}
\toprule
   &  SISDR (dB) & BSS SDR (dB) & Params (M) & MS (MB) & WM (KB) & MOps/inf. & Latency (ms/inf.) & Energy (mJ/inf.) & GPUH\\
\midrule

Baseline (FP32)      &    11.99        &      12.77      &   0.97   &  \textcolor{red}{3.70}

&         \textcolor{blue}{26.0}            &    \textcolor{red}{1.94}       &   \textcolor{red}{12.52$^*$}               &   6.76$^*$      &  14 \\

Pruned (FP32)  &    11.99        &     12.78     &  0.52   & \textcolor{red}{1.97} 
&       \textcolor{blue}{18.8}              &     \textcolor{blue}{1.04}      &     \textcolor{blue}{6.71$^*$}            &           3.62$^*$      & 72\\
\midrule
Pruned (INT8) 1   &    11.80        &     12.69       &   0.61  & \textcolor{red}{0.58}
&         \textcolor{blue}{5.1}          &     \textcolor{blue}{1.22}      &       \textcolor{blue}{7.87}          &         4.25        & 61\\
Pruned (INT8) 2   &     11.47      &     12.22       &   0.33  & \textcolor{blue}{0.31} 
&        \textcolor{blue}{3.7}             & \textcolor{blue}{0.66}      &   \textcolor{blue}{4.26}     &             2.30    & 144\\
Pruned Skip RNN (INT8)   &      11.42       &      12.07     &   0.46   &     \textcolor{blue}{0.43} 
&        \textcolor{blue}{4.67}        &    \textcolor{blue}{0.37$^\dagger$}       &      \textcolor{blue}{2.39$^\dagger$}           &  1.29$^\dagger$ & 275 \\
\midrule
Erdogan et al.~\cite{erdogan2015phase}           &    -     &     13.36         &   0.96
& \textcolor{red}{3.65}
&  \textcolor{blue}{25.9}  &  \textcolor{red}{1.92}  &  \textcolor{red}{12.39}  &  6.69 &   -  \\
Wilson et al.~\cite{wilson2018exploring}         &    -         &    14.60          &     65 
&  \textcolor{red}{247.96}
& \textcolor{red}{4472.6}        &      \textcolor{red}{130} &      \textcolor{red}{839$^*$}           &     453            & 18360 \\
\bottomrule
\end{tabular}}
\caption{Model performance on CHiME2 development set and STM32F746VE, which was measured to run at 155MOps/s while drawing 0.54W. The symbol $^*$ denotes best-case estimates because the underlying models are in floating point and measurements were taken for integer arithmetic. The symbol $^\dagger$ denotes average performance, reflecting the stochasticity of the skip RNN model. Blue (red) denotes measurements that pass (violate) one of the constraints in Table \ref{table:constraints}. Measurements do not include cost of STFT or Mel transforms.\vspace{-2em}}
\label{table:optimizations}
\end{table*}

The MS is the total number of parameters in all layers, multiplied by the datatype of each matrix. The number of operations required per inference also depends on the number of parameters, since (almost) all of the operations performed in our model are matrix-vector multiplications, which require 2 ops (multiply and add) per parameter. Although the operation count is independent of model quantization, the throughput that can be achieved on real HW is much higher with lower precision integer data types.
Therefore, to reduce overall latency, we employ two optimizations: 1) pruning to reduce operations, and 2) 
weight/activation quantization, which reduces MS and 
enables deployment with low-precision integer arithmetic \cite{jacob2018}.



\subsection{Structural Pruning}
\label{sec:structural pruning}
Pruning is a well established network optimization \cite{lecun1990optimal,mozer1989skeletonization}. We specifically use structured pruning because it leads to direct benefits in both model size and throughput \cite{wen2016learning}. This is in contrast to random pruning, which is harder to exploit on real HW, unless the sparsity is very high.
We begin by grouping weights in $\theta$ into a set $\Gamma$, where $w_g \in \Gamma$ denotes the set of weights in a particular group, stacked into a vector. The organization of the groups defines the kind of structures we can learn. For FC layers, we group weights by the neuron they are connected to in the previous layer. For LSTM layers, we group weights by the element of $h^t$ to which they are connected \cite{wen2018learning}. We assign a binary mask $r_g = \mathbbm{1}\left( \Vert w_g \Vert_2 - \tau_k \geq 0 \right)$ to each group of weights in layer $k$, where $\mathbbm{1}\left[ \cdot \right]$ denotes the indicator function and $\tau_k \geq 0$ is a learnable threshold. Let $P = \left \lbrace r_g, 1 \leq g \leq \vert \Gamma \vert \right \rbrace$ be the set of pruning masks and $\theta \odot P$ be shorthand for the set of model weights with each weight multiplied by its corresponding binary mask. We then modify the learning objective in \eqref{eq:empirical risk} to penalize the energy of non-zero weights:\vspace{-1em}
\begin{align}\label{eq:pruned empirical risk}
    \min_{\theta,\left \lbrace \tau_k, 1 \leq k \leq K \right \rbrace} {L}(\theta \odot {P}) + \lambda \sum_{g=1}^{\vert \Gamma \vert} r_g \left \Vert w_g \right \Vert_2 
\end{align}
where $\lambda$ is a hyperparameter controlling the degree of pruning and $K$ is the number of layers. To differentiate the indicator function, we approximate it by the sigmoid function in the backward pass \cite{stamoulis2020}. The pruning approach described above is unique amongst the pruning literature and is particularly suited for our particular task. We adopt the structural grouping of LSTM weights from \cite{wen2018learning}, but we improve upon the manually selected pruning thresholds in \cite{wen2018learning} by learning them directly. The result is that we do not need to perform a hyperparameter search for $\left \lbrace \tau_k, 1 \leq k \leq K \right \rbrace$, which can be prohibitively costly since SE RNNs take roughly 14 GPUH to train and the hyperparameter space grows exponentially with $K$.

\subsection{Quantization}
\label{sec:quantization}
Let $w \in \mathbb{R}$ denote a real-valued (floating point) value and $Q_{\alpha,\beta}(w)$ its quantized value, where the quantization is performed uniformly over the range $(\alpha,\beta)$ with $2^{\text{bits}}-1$ levels, i.e. $Q_{\alpha,\beta}(w) = \zeta \text{round}\left( (\text{clip}\left(w,\alpha,\beta\right) - \alpha) / \zeta \right) + \alpha$,
where $\alpha < \beta$ and $\zeta = (2^{\text{bits}-1})(\beta - \alpha)$. For brevity, we omit $\alpha$ and $\beta$ subscripts moving forward. We adopt a standard approach of making the model resilient to quantized tensors by performing training-aware quantization \cite{jacob2018}. This exposes the model outputs to quantization noise, while still allowing the model to backpropagate on real-valued weights. Concretely, 
\eqref{eq:pruned empirical risk} becomes
\begin{align}\label{eq:pruned quantized empirical risk}
    \min_{\substack{\theta,\Omega \\ \left \lbrace \tau_k, 1 \leq k \leq K \right \rbrace }} {L}_{Q}\left(Q\left(\theta \odot P \right)\right) + \lambda \sum_{g=1}^{\vert \Gamma \vert} r_g \left \Vert Q \left( w_g \right) \right \Vert_2 
\end{align}
where $\Omega$ is the set of quantization parameters for all weights and activations, $Q\left(\theta \odot P\right)$ represents the fact that the masked network weights are quantized, and ${L}_{Q}$ denotes that activations are quantized. During backpropagation, the $\text{round}(\cdot)$ operation is ignored \cite{jacob2018}. We quantize the weights, activations and model input to 8 bits;
the mask itself, is quantized to 16 bits. 

\subsection{Skip RNN Cells}
Finally, we evaluated the skip RNN approach~\cite{campos2018}, which can be considered a form of dynamic temporal pruning. A binary neuron $g^t \in \lbrace 0,1 \rbrace$ is introduced that acts as a {state update gate} on the candidate LSTM states $\tilde{s}$, representing both $c^t$ and $h^t$ from \eqref{eq:lstm 5}:
\begin{align}
g^t &= \text{round}\left(\tilde{g}^t \right) \label{eq:skip 1}, \; \;  s^t = g^t \tilde{s}^t + (1-g^t)  s^{t-1}
\end{align}
where $\tilde{g}^t$ is the update probability, updated using
\begin{align}
\Delta \tilde{g}^t &= \sigma \left( W_b c_{*}^{t-1} + b_b \right) \label{eq: skip 2} \\
\tilde{g}^{t+1} &= g^t  \Delta \tilde{g}^t + (1-g^t)  (\tilde g^t + min(\Delta \tilde g^t, 1 - \tilde g^t))
\end{align}
where $c_{*}^{t-1}$ is the state of the final LSTM layer. Whenever a state update is skipped, the state update probability $\tilde{g}^t$ is incremented by $\Delta \tilde{g}^t$ until $\tilde{g}^t$ is high enough that an update occurs, in which case  $\tilde{g}^{t+1}$ becomes $ \Delta \tilde{g}^t$.
Because $\tilde{g}^t$ is fixed when the LSTMs are not updating, \eqref{eq: skip 2} only needs to be calculated when the LSTMs are updated.

In practice, this method of skipping updates 
performs well on training and evaluation metrics, but produced audio artifacts as the mask itself is not updated when the LSTM skips. To remedy this, two exponential moving averages (EMAs) were introduced to smooth the model in time. First, a context vector, $c_x^t = 0.9 c_x^{t-1} + 0.1 \left( W_c x^t + b_c \right)$, is calculated as an EMA projected from the input spectrogram frame, $x_t$, and concatenated to the LSTM output. Second, an EMA is applied to the masks, $m^t$, to calculate a smoothed mask $\tilde{m}_t = 0.15 \tilde{m}^{t-1} + 0.85 m^t$.


\label{sec:skip rnn}
\section{Experimental Results}
\renewcommand{\thefootnote}{\fnsymbol{footnote}}

\begin{table}
\resizebox{\columnwidth}{!}{
\begin{tabular}{llllllll}
\hline
\multicolumn{1}{c}{SDR {[}dB{]}} & \multicolumn{7}{c}{Input SNR {[}dB{]}}                \\
\multicolumn{1}{c}{}                              & -6    & -3    & 0     & 3     & 6     & 9     & Avg.  \\ \hline
Baseline (FP32)                                         & 10.01  & 11.54 & 13.08 & 14.23 & 15.85 & 17.46 & 13.70 \\
Pruned  (FP32)                                  &  10.07  & 11.59 & 13.10 & 14.31 &  15.89  & 17.50 & 13.74 \\ \hline
Pruned (INT8) 1                    &  9.82  & 11.37 & 12.91 & 14.20 & 15.74 & 17.44 &  13.58\\ 
Pruned (INT8) 2                                   &  9.33  &  10.91 & 12.46 & 13.79 & 15.46 & 17.16  & 13.18 \\ 
Pruned Skip RNN (INT8)                                   &   9.13  & 10.70  & 12.27  & 13.54 &  15.20  &  16.93 & 12.96 \\ \hline
Erdogan et al.~\cite{erdogan2015phase}                                              & -  & -  & - & - & - & - & 14.14 \\
Wilson et al.~\cite{wilson2018exploring}                                            & 12.17  & 13.44  & 14.70 & 15.83 & 17.30 & 18.78 & 15.37 \\
\bottomrule
\end{tabular}}
\caption{Model performance evaluated on CHiME2 test set.
}
\vspace{-2em}
\label{table:bss_eval}
\end{table}

For all experiments, we optimize the objective using stochastic gradient descent in Tensorflow. 
We use a 32ms frame, 16ms hop size, and 16KHz sample rate
for our baseline, pruning, and quantization experiments. For skip RNN experiments, we use frame and hop sizes of 25ms and 6.25ms, respectively.
All methods are trained and evaluated with the CHiME2 WSJ0 dataset \cite{vincent2013second},
which contains 7138 training, 2560 development, and 1980 test utterances, respectively.
All three subsets include utterances with signal-to-noise ratios (SNRs) in the range -6 to 9dB. Noise data consists of highly non-stationary interference sources recorded in living room settings including vacuum cleaners, television and children. While the dataset is provided in binaural stereo, we pre-process by summing over the channel dimension to obtain a monaural input and target, in contrast to \cite{wilson2018exploring}, which uses the full binaural input to predict a binaural mask. For final objective evaluations, we use the signal-to-distortion ratio (SDR) \cite{vincent2006performance}. During training, however, we use the simpler scale-invariant signal-to-distortion ratio (SISDR) since it is computationally less expensive and correlates well to SDR \cite{le2019sdr}.  

\subsection{Baseline Model}
We begin by confirming that our baseline SE model is competitive with the state of the art. 
Our baseline achieves 12.77dB SDR on the CHiME2 development set (Table~\ref{table:optimizations}) and 13.70dB SDR on the test set (Table~\ref{table:bss_eval}), which is comparable to \cite{weninger2014discriminatively,erdogan2015phase}.
\vspace{-0.5em}
\subsection{Structured Pruning and Quantization}
\begin{figure}
\centering
\includegraphics[width=\linewidth]{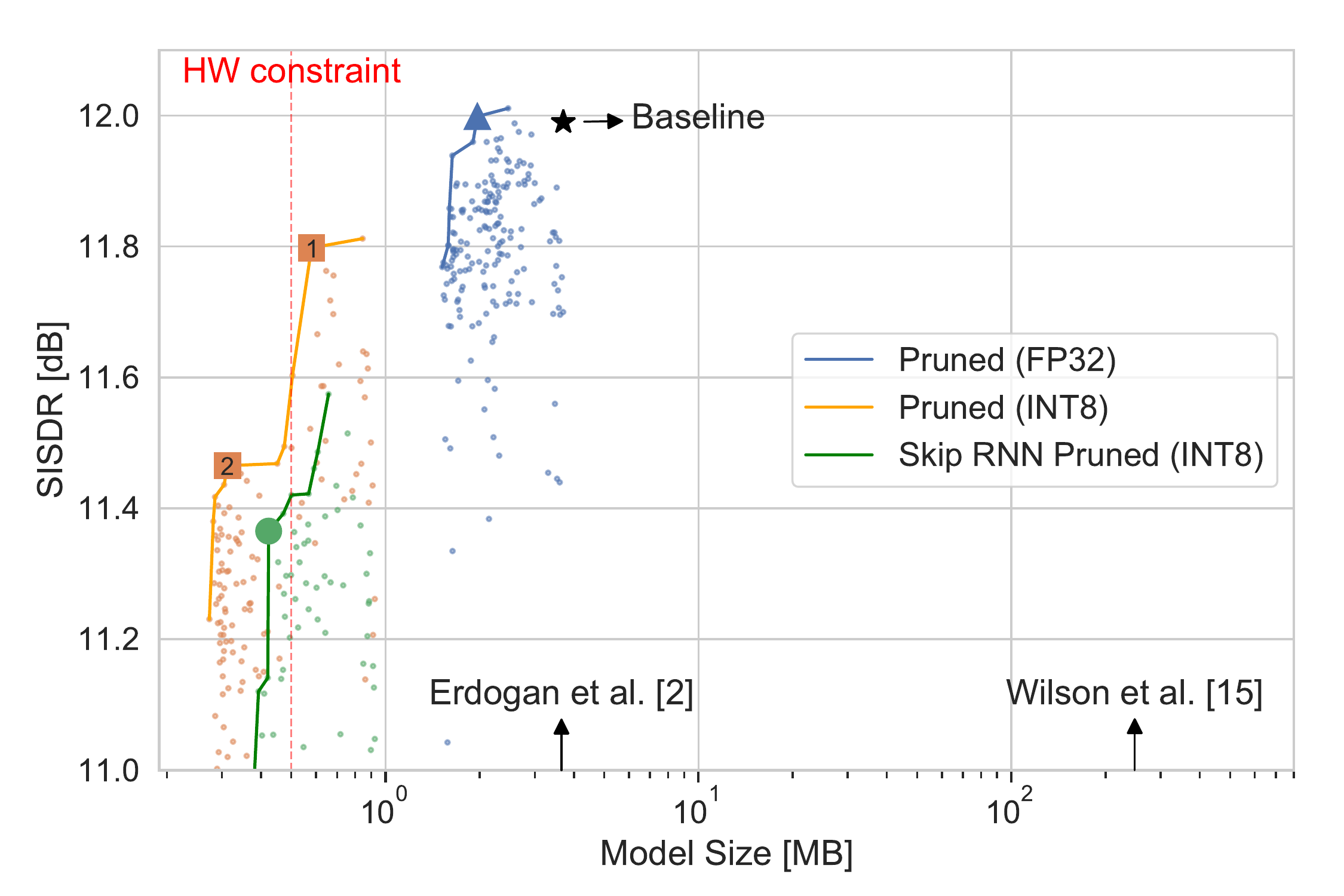}
\vspace{-2em}
\caption{MS vs. SISDR. Each point represents a model checkpoint and lines represent a Pareto front.
}
\label{fig: size_vs_sisnr}
\end{figure}

Next, 
we examine the effect of structural pruning and quantization on the baseline model. 
In all cases, we set $\lambda = 10^{-9}$.
The trade-off between model size and performance is illustrated in Fig.~\ref{fig: size_vs_sisnr}, where each point represents a snapshot during the optimization process. We plot the pareto frontier of SISDR values with respect to the MS.
Our experiments show that structural pruning can achieve
a 47\% pruned model with identical performance to the baseline. Moreover, with both pruning and quantization in effect, a 37\% pruned model achieves around 0.2dB reduction in SISDR (Pruned (INT8) 1), and a 66\% pruned model shows around 0.5dB decay (Pruned (INT8) 2). Table \ref{table:bss_eval} shows SDR evaluations of our models on the CHiME2 test set.

Our optimized models
achieve latency suitable for a smaller frame processing time (hop size) in the audio pipeline. However, a small hop size leads to increased inference frequency and energy consumption. So, to address this challenge, we apply the skip RNN architecture on top of our compressed model. 
Results for ``Pruned Skip RNN (INT8)'' show 12.07dB SDR on the CHiME2 development set (Table~\ref{table:optimizations}) and 12.96dB SDR on the test set (Table~\ref{table:bss_eval}).  Although the skip RNN requires more inferences per second, the skip rate of 63\% results in reduced average energy comsumtion per inference compared to Pruned (INT8) 2.

Finally, Table \ref{table:optimizations} details each of the models described. 
Although the models in \cite{erdogan2015phase,wilson2018exploring} achieve slightly better SISDR/SDR performance, their MS, WM, and MOps/inf severely violate HA HW constraints. In contrast, Pruned (INT8) model 2 and Pruned Skip RNN (INT8) can be deployed on a real HA MCU and deliver significant SE capabilities. In contrast to \cite{erdogan2015phase,wilson2018exploring}, our models achieve compute latency in the 2.39-6.71ms range, satisfying the 10ms requirement. Moreover, the proposed models consume significantly less energy per inference than \cite{erdogan2015phase,wilson2018exploring}, leading to better HA battery life.

\subsection{Perceptual Evaluation}
\vspace{-0.5em}
Human perception of audio quality is highly subjective and 
does not always correlate with objective metrics. 
Therefore, to understand real-world performance, we conducted perceptual studies to get subjective feedback on the quality of the optimized model compared to our baseline.
We conducted surveys for both Pruned (INT8) models (Table~\ref{table:optimizations}), each consisting of disjoint sets of 50 participants. Two samples were randomly chosen from each of the 6 SNR levels of the CHiME2 evaluation set, for a total of 12 sample utterances. Each participant was randomly presented with paired comparisons of the original and processed utterance for both the baseline and pruned \& quantized models, resulting in 24 paired comparisons per participant. Given the prompt \emph{“Considering both clarity and quality of speech, which recording do you prefer?”}, the participants rated comparative preference on a 7-point Likert scale \cite{likert1932} ranging from ``Strongly prefer unprocessed’' to ``Strongly prefer enhanced'', with ``No preference'' as a midpoint.

\begin{figure}
\centering
\includegraphics[width=\linewidth]{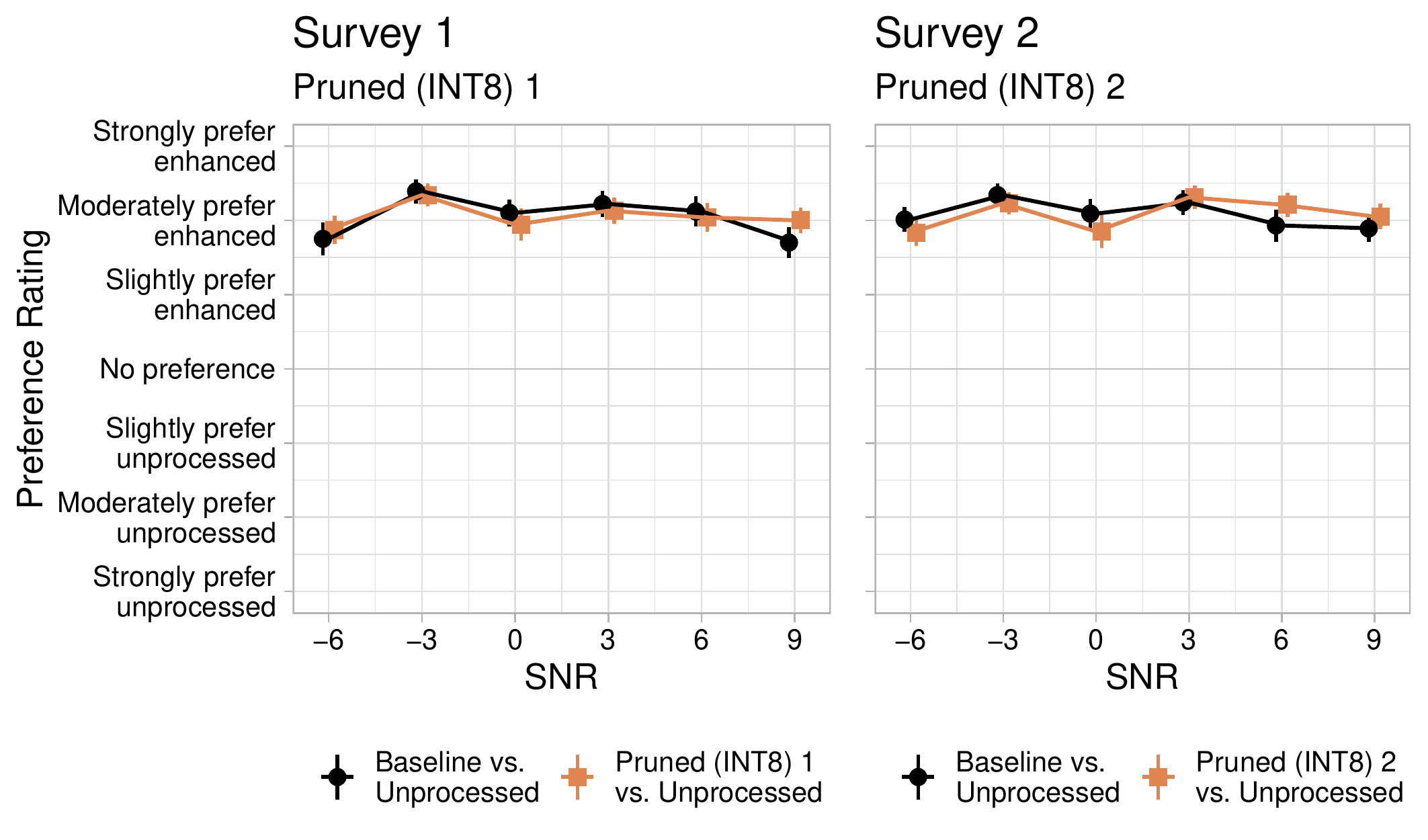}
\vspace{-2em}
\caption{Preference of perceptual study participants for enhanced audio vs. unprocessed audio for both uncompressed and pruned \& quantized models across input SNR's. Left is Pruned (INT8) 1 and right is Pruned (INT8) 2.
}
\label{fig:percpetual evaluation}
\end{figure}

The results in Fig.~\ref{fig:percpetual evaluation}, show that participants expressed a ``moderate preference'' on average for the enhanced audio. We note this compares favorably with industry-standard approaches to improving HA speech-in-noise performance, where participants in a similar study expressed a ``slight preference'' for directionally processed audio compared to unprocessed \cite{vaisberg2020}. We compare preference for the uncompressed (baseline) and the compressed (pruned and quantized) models to the original unprocessed utterances using a Wilcoxon signed rank test \cite{wilcoxon1945} and find no statistical difference between ratings across SNRs (Survey 1: Z = 0.09, p = 0.92; Survey 2:  Z = 0.19, p = 0.85), indicating that participants prefer the enhanced audio, independently of whether it was produced by the baseline or optimized model. 





\vspace{-1em}
\section{Conclusions}
Neural speech enhancement is a key technology for future HA products. However, the latency and power consumption constraints for tiny battery-powered HW is extremely hard to meet due to the large NNs required to achieve satisfactory audio performance.
In this work, we applied structural pruning and integer quantization of inputs, weights, and activations to reduce model size by 11.9$\times$, compared to the baseline. We also applied skip RNN techniques to further reduce the ops per inference by 1.78$\times$ compared to our smallest compressed model. Our optimized models demonstrate negligible degradation in objective (SISDR) metrics and no statistical difference in subjective human perceptual evaluation. While our baseline model has a computational latency of 12.52ms on our target HW platform, the optimized implementation achieves 4.26ms, which is more than sufficient to meet the 10ms compute latency target.


\bibliographystyle{IEEEtran}
\bibliography{mybib}


\end{document}